\documentclass[12pt]{article}
\usepackage{amsmath,amssymb,exscale}  
\input epsf

\def\gappeq{\mathrel{\rlap {\raise.5ex\hbox{$>$}}
{\lower.5ex\hbox{$\sim$}}}}
\def\lappeq{\mathrel{\rlap{\raise.5ex\hbox{$<$}}
{\lower.5ex\hbox{$\sim$}}}}

\begin{document}
\topmargin -1.0cm
\oddsidemargin -0.8cm
\evensidemargin -0.8cm
\pagestyle{empty}
\begin{flushright}
UNIL-IPT-00-28\\
December 2000
\end{flushright}
\vspace*{5mm}

\begin{center}

{\Large\bf A Warped Supersymmetric Standard Model}\\
\vspace{1.0cm}

{\large Tony Gherghetta$^a$ and 
Alex Pomarol$^{b}$}\\
\vspace{.6cm}
{\it {$^{a}$IPT, University of Lausanne, 
CH-1015 Lausanne, Switzerland}}\\
{\it {$^{b}$ IFAE, Universitat Aut{\`o}noma de Barcelona, 
E-08193 Barcelona, Spain}}\\
\vspace{.4cm}
\end{center}

\vspace{1cm}
\begin{abstract}

We study the breaking of supersymmetry in five-dimensional 
(5d) warped spaces, using the Randall-Sundrum model 
as a prototype. In particular, we present a 
supersymmetry-breaking mechanism which has a geometrical origin, 
and consists of imposing different boundary conditions between
the fermions and bosons living in the 5d bulk.
The scale of supersymmetry breaking is exponentially
small due to the warp factor of the AdS metric.
We apply this mechanism to a  supersymmetric standard model where
supersymmetry breaking is transmitted through the AdS bulk to 
matter fields confined on the Planck-brane.
This leads to a  predictable superparticle mass spectrum  
where the gravitino mass is $10^{-3}$eV and scalar particles 
receive masses at the one-loop level via bulk gauge interactions. 
We calculate the mass spectrum 
in full detail using the 5d AdS propagators.
The AdS/CFT correspondence suggests that our 5d warped model is dual to the
ordinary 4d MSSM with a strongly coupled CFT sector responsible for the
breaking of supersymmetry.

\end{abstract}

\vfill
\begin{flushleft}
\end{flushleft}
\eject
\pagestyle{empty}
\setcounter{page}{1}
\setcounter{footnote}{0}
\pagestyle{plain}

%============================================================

\section{Introduction}

The standard model is believed to 
be  an effective theory valid up to  some energy scale
near the electroweak scale. 
What lies beyond the standard model has been the subject of active research.
Among the possible candidates, there are technicolor theories, supersymmetry,
and, recently, extra dimensions~\cite{add}. 

Extra dimensions and supersymmetry present an  additional
motivation.  They could  be an important ingredient
in the  underlying theory that  
includes a quantum description of gravity, 
and in particular for string theory they play a
crucial role.
A particularly interesting extra dimension scenario
is the Randall-Sundrum model~\cite{rs}. 
In this model the extra dimension is 
compactified in a slice of anti-deSitter 
(AdS) space, and, as a consequence,  the electroweak 
scale is generated by an exponential warp factor in the metric.
This  model can be supersymmetrized~\cite{gp,Alten,susyRS}
providing  an interesting alternative to the minimal
supersymmetric standard model (MSSM), and 
a possible connection to string theories~\cite{stringRS}.

In this article we want to continue the study of supersymmetric extensions
of the standard model living in five dimensions where the extra dimension
is compactified as in the  Randall-Sundrum model~\cite{gp,log}.
In particular, we want to study supersymmetry breaking.
A warped extra dimension allows for new ways of breaking supersymmetry.
The particular mechanism that we will consider here 
consists of imposing different boundary conditions between the 
fermions and bosons.
This supersymmetry breaking mechanism has been previously 
studied in flat space but not in warped spaces. 
In warped space this way of 
breaking supersymmetry leads to novel phenomenological consequences.
For example, the radius of compactification does
not need to be large (TeV$^{-1}$) as in the case of flat space \cite{anto}.
Therefore it can be consistent with a large cut-off scale that is
related with the Planck scale, $M_P$, or grand unified 
theories~\cite{gauge,log}.
As we will show, the
scale of supersymmetry breaking can be  very low ($\sim$ TeV) and
this implies a superlight gravitino $m_{3/2}\sim 10^{-3}$ eV. 
Scalars are massless at tree-level and get masses at the one-loop level.
We will study in detail a ``warped'' version of the MSSM,
where gravity and  gauge bosons 
live in the five-dimensional (5d) AdS bulk, 
while matter fields are located on one 
of the boundaries, the Planck-brane.
In this warped MSSM
the squark and slepton masses arise at one-loop  from  the gauge interactions
and are therefore naturally flavor independent.
One of the most interesting properties of the model
is its  predictivity of the  low-energy mass spectrum.
We will calculate it here in full detail.
Although we present the calculation for a particular extension of the 
standard model, the calculation of quantum 
effects in warped spaces that we present here is much more general 
and can also be useful for other scenarios.

Another important motivation for the study of the MSSM  
in a slice of AdS arises from the AdS/CFT correspondence~\cite{maldacena}.
This conjecture suggests that these 5d models have a strongly-coupled 
4d dual~\cite{gu,gkr,apr,rz}.
Therefore, the study of the weakly coupled 5d gravity theory here will
be helpful in understanding supersymmetric 4d theories with 
a strongly coupled sector. 
We will comment later on this duality.

In section~2 we introduce the Randall-Sundrum compactification
and its supersymmetric version. In particular we
also analyze the gravitino Kaluza-Klein decomposition since it is 
the only field not considered in Ref.~\cite{gp} (see also~\cite{oda}).
In section~3, we present the supersymmetry-breaking 
mechanism, which is based on imposing ``twisted'' boundary conditions 
for fermions in the bulk, and comment on the differences compared
with the case of a flat extra dimension. 
In section~4 we introduce a version of the MSSM 
living in a slice of AdS and calculate the sparticle mass spectrum,
at tree-level and at one-loop level.
We will also comment on the holographic interpretation of the model.
Our concluding remarks appear in Section~5. 
Finally in the Appendix, we present a detailed calculation
of the 5d propagators in a slice of AdS.

\section{The Warped Supersymmetric Brane-World}

We will consider the scenario of Ref.~\cite{rs}, which is based on 
a nonfactorizable 5d geometry. 
The fifth dimension
$y$ is compactified on an orbifold, $S^1/Z_2$ of radius $R$, with 
$0 \leq y \leq \pi R$. The boundary of the 5d spacetime consists of
two 3-branes located at the orbifold fixed points 
$y^\ast =0$ and $y^\ast=\pi R$.
This configuration with the 5d metric solution~\cite{rs}
\begin{equation}
   ds^2 = e^{-2\sigma} \eta_{\mu\nu} dx^\mu dx^\nu + dy^2\equiv 
         g_{MN} dx^M dx^N~,
\label{metric}
\end{equation}
is a slice of AdS space, 
where $\sigma = k|y|$ and $1/k$ is the AdS curvature radius.
The 5d coordinates are labelled by capital Latin letters, $M=(\mu,5)$
where $\mu=0,\dots,3$.
The complete supergravity action for this configuration 
is obtained by including the gravitino and graviphoton together with the
graviton \cite{gp,Alten,susyRS}. 
However, for the discussion of supersymmetry breaking 
it will suffice to only consider the additional gravitino kinetic 
and mass terms, which are given by~\cite{gp}
\begin{eqnarray}
\label{action}
    S&=&S_5 + S_{(0)}+S_{(\pi R)}~,\nonumber \\
    S_5 &=& \int d^4x \int dy \sqrt{-g}\, \Bigg[ -\frac{1}{2} M_5^3
    \Big({\cal R}+i\bar\Psi^i_M\gamma^{MNP}D_N\Psi_{P}^i
    -i\frac{3}{2}\sigma^\prime \bar\Psi^i_M\gamma^{MN}(\sigma_3)^{ij}
    \Psi_{N}^j \Big) - \Lambda \Bigg]~, \nonumber \\
    S_{(y^\ast)} &=& \int d^4x\sqrt{-g_4} 
      \left[ {\cal L}_{(y^\ast)} -\Lambda_{(y^\ast)}  \right]~,
\end{eqnarray}
where $g_4$ is the induced metric on the 3-brane located at $y^\ast$, and
$\gamma^{M_1 M_2\dots M_n}=\frac{1}{n!}\gamma^{[M_1}\gamma^{M_2}
\dots\gamma^{M_n]}$ is the antisymmetrized product of gamma matrices. 
We have defined $\sigma^\prime=d\sigma/dy$.
Supersymmetry automatically ensures the bulk/boundary conditions  
$\Lambda_{(0)}=-\Lambda_{(\pi R)}= -\Lambda/k$.
The action contains the 5d Planck scale $M_5$, 
the 5d Ricci scalar $\cal R$, two symplectic-Majorana gravitinos, 
$\Psi_M^i$ $(i=1,2)$, and a bulk cosmological constant $\Lambda$. 
At $y^\ast=0$ the effective 4d mass scale
is of order of the Planck scale, $M_P^2\simeq M^3_5/k$, 
and we will refer to the brane there as the 
Planck-brane. Similarly, at $y^\ast = \pi R$ the effective mass scale is 
of order $M_P e^{-\pi kR}$, which is
the TeV scale for $kR\simeq 11$. Consequently the 3-brane located 
there will be referred to as the TeV-brane.
The index $i$ labels the fundamental representation of the SU(2)$_R$ 
automorphism
group of the $N=1$ supersymmetry algebra in five dimensions. 
The gravitino supersymmetry
transformation is given by~\cite{gp}
\begin{equation}
        \delta\Psi_M^i = D_M\eta^i +\frac{\sigma^\prime}{2} 
        \gamma_M (\sigma_3)^{ij} \eta^j~,
\label{susyt}
\end{equation}
where the symplectic-Majorana spinor $\eta^i$ is the 
5d supersymmetry parameter.

Similarly, gauge bosons and matter can be added to the bulk 
\cite{gw,dhr,gauge,smbulk,gp}. 
In a 5d supersymmetric theory they 
form part of vector supermultiplets and hypermultiplets.
The behavior of these supermultiplets in the background of 
Eq.~(\ref{metric}) was considered in Ref.~\cite{gp},
where the Kaluza-Klein mass spectrum was also derived. 
Only the analysis of the gravitino field was not presented 
in Ref.~\cite{gp}. For this reason, we will present below the 
Kaluza-Klein decomposition of the gravitino.
This will also help
to show how the superHiggs mechanism operates level-by-level
in the Kaluza-Klein modes, and will help to
better understand the 
supersymmetry breaking mechanism presented in the next section.

\subsection{Kaluza-Klein decomposition of the gravitino and the 
superHiggs mechanism level by level}    
\label{superH}

Let us start by 
decomposing the 5d gravitino, $\Psi_M$, and the 5d supersymmetry 
parameter, $\eta$, into 4d Kaluza-Klein fields
\begin{eqnarray}
    \Psi_{\mu\, L,R}(x^\mu,y) &=& \sum_{n=0}^\infty
    \psi_{\mu\, L,R}^{(n)}(x^\mu)\, f^{(n)}_{L,R}(y)~,\nonumber\\
    \Psi_{5\, L,R}(x^\mu,y)&=& 
\sum_{n=0}^\infty \psi_{5\, L,R}^{(n)}(x^\mu) f^{(n)}_{5\, L,R}(y)~,\nonumber\\
    \eta_{L,R}(x^\mu,y) &=& \sum_{n=0}^\infty \eta_{L,R}^{(n)}(x^\mu)
      f^{(n)}_{L,R}(y)~.
\label{KKdecom}
\end{eqnarray}
We have dropped the SU(2)$_R$ index 
$i$, since we need only consider the $i=1$ component (the $i=2$
component is simply obtained from  the
symplectic-Majorana condition).
We have also defined $\gamma_5\Psi_{L,R}=\pm\Psi_{L,R}$.
It is important to note that 
we have chosen the $y$-dependent 
wavefunction of the supersymmetry parameter
$\eta$ to be the same as that for the Kaluza-Klein gravitinos.

\subsubsection{Kaluza-Klein modes $n\not=0$}

The supersymmetry  transformation Eq.~(\ref{susyt}) for $i=1$ gives
\begin{eqnarray}
\label{susytrans}
      \delta \Psi_\mu&=&\partial_\mu\eta+\sigma^\prime\gamma_\mu
       \left(\frac{1-\gamma_5}{2}\right) \eta~,\nonumber\\
      \delta \Psi_5&=&\partial_5 \eta+\sigma^\prime
      \frac{\gamma_5}{2} \eta~.
\label{susytrans5}
\end{eqnarray}
Substituting Eq.~(\ref{KKdecom}) into Eq.~(\ref{susytrans}) and 
projecting out the $n$th-mode~\footnote{
This corresponds to multiplying each side of Eq.~(\ref{susytrans})
by $f^{(n)}_{L,R}$, integrating over $y$, and using the gravitino 
orthogonality condition 
$\int dy\, e^{-\sigma} f^{(n)}_{L,R} f^{(m)}_{L,R} = \delta_{nm}$.},
we find that 
the supersymmetry transformation for the $n$th Kaluza-Klein gravitino mode 
is given by
\begin{eqnarray}
     \delta \psi_{\mu\, L}^{(n)} &=& \partial_\mu \eta_L^{(n)}
     +\tilde\gamma_\mu \sum_{k=0}^\infty a_{nk}\eta^{(k)}_R~, \\
     \delta \psi_{\mu\, R}^{(n)}&=&\partial_\mu \eta_R^{(n)}~,
\end{eqnarray}
where $\tilde\gamma_\mu$ is the 4d Minkowski gamma matrix and the 
coefficients $a_{nk}$
are given by
\begin{equation}
     a_{nk}\equiv\int 
      dy\, e^{-2\sigma} \sigma^\prime f^{(n)}_L(y) f^{(k)}_R(y)~. 
\end{equation}
The coefficients $a_{nk}$ imply that the 
supersymmetry transformation of $\psi_{\mu L}^{(n)}$ at level $n$,
depends nontrivially on the complete tower of
Kaluza-Klein parameters $\eta_R^{(k)}$. This
effect is entirely due to the fact that the bulk is AdS.
Let us now impose the following relation for the 
wavefunctions of   $\Psi_5$:
\begin{equation}
     f^{(n)}_{5\, {L,R}}=\frac{1}{m_n}
     \left( \pm \partial_5 + \frac{1}{2}\sigma^\prime\right) f^{(n)}_{L,R}~,
\label{condi}
\end{equation}
where $m_n$ is the 4d mass of the gravitino Kaluza-Klein mode $n$, 
which will be derived below.
The condition (\ref{condi})  allows us to write a simple expression for the 
supersymmetry transformation of Kaluza-Klein modes $\psi_5^{(n)}$
\begin{eqnarray}
    \label{varL}
\delta \psi_{5\, L}^{(n)}&=&m_n \eta_L^{(n)}~, \\
    \delta \psi_{5\, R}^{(n)}&=&-m_n\eta_R^{(n)} \label{varR}~.
\end{eqnarray}
This shows that the $n$th Kaluza-Klein mode of the 5th component
of the gravitino transforms as a Goldstino of the 
$\eta^{(n)}$ supersymmetry transformation and that 
these $N=2$ supersymmetries are non-linearly realized.
We can now see that the redefined gravitinos
\begin{eqnarray}
    \widetilde\psi_{\mu\, L}^{(n)}&\equiv&m_n\psi_{\mu\, L}^{(n)}-
    \partial_\mu\psi^{(n)}_{5\, L} +
    m_n \tilde\gamma_\mu \sum_{k=0}^\infty
    a_{nk} \frac{\psi^{(k)}_{5\, R}}{m_k}\nonumber~,\\
    \widetilde\psi_{\mu\, R}^{(n)}&\equiv&m_n\psi_{\mu\, R}^{(n)}+\partial_\mu
    \psi^{(n)}_{5\, R}~,
\label{redef}
\end{eqnarray}
are invariant under supersymmetry transformations, and therefore correspond to
the physical fields. On the contrary,  the fields 
$\psi_5^{(n)}$ 
are gauge dependent and can be eliminated.
This is the superHiggs mechanism.  The $\psi_5^{(n)}$ are eaten
by the gravitino $\psi_{\mu}^{(n)}$ to become massive.

Let us now turn to
the Rarita-Schwinger equation for the bulk
gravitino, which in the AdS background reads
\begin{equation}
   \gamma^{MNP}D_N\Psi_P-\frac{3}{2}\sigma^\prime\gamma^{MP}
   \Psi_P=0~.
\label{rari}
\end{equation}
Using the redefined gravitino fields (\ref{redef}), the equation of motion
(\ref{rari}) simplifies to
\begin{equation}
  \gamma^{\mu\nu\rho}\partial_\nu\widetilde \psi^{(n)}_{\rho\, L,R}
   -m_n\gamma^{\mu\rho} \widetilde \psi^{(n)}_{\rho\, R,L}=0~,
\end{equation}
which represents the 4d massive Rarita-Schwinger equation for
the spin 3/2 field $\widetilde\psi_\mu^{(n)}$, and
where the $y$-dependent Kaluza-Klein wavefunctions satisfy
\begin{eqnarray}
    \left(\partial_5+\frac{1}{2}\sigma^\prime\right) f^{(n)}_L
    &=&m_ne^\sigma f_R^{(n)}\label{gravi1}~,\\
    \left( \partial_5-\frac{5}{2}\sigma^\prime\right) f^{(n)}_R
    &=&-m_ne^\sigma f_L^{(n)}\label{gravi2}~.
\end{eqnarray}
One can see that the dependence on $\psi_5^{(n)}$ has dropped out and 
the equation of motion depends, as expected,
 only on $\widetilde\psi_\mu^{(n)}$.
The solutions of Eqs.~(\ref{gravi1}) and (\ref{gravi2}) are a special
case of the general solution appearing in Ref.~\cite{gp}. In fact, defining 
$\widehat f^{(n)}_{L,R} = e^{-\sigma} f^{(n)}_{L,R}$
one can see that $\widehat f^{(n)}_{L,R}$ corresponds to the wavefunction 
of a ``hatted'' fermion of mass $m=3\sigma^\prime/2$ 
defined in Ref.~\cite{gp}.
Thus, using the results in
Ref.~\cite{gp} and the fact that $\Psi_{\mu\, L}$
($\Psi_{\mu\, R}$) are defined even (odd) under the 
$Z_2$-parity,
we obtain the $y$-dependent gravitino wavefunctions
\begin{eqnarray}
   f^{(n)}_{L}&=&\frac{1}{N_n}e^{\frac{3}{2}\sigma}
   \left[J_{2}(\frac{m_n}{k}e^\sigma)+b(m_n)
   Y_{2}(\frac{m_n}{k}e^\sigma)\right]\, ,\\
   f^{(n)}_{R}&=&\frac{\sigma^\prime}{kN_n}e^{\frac{3}{2}\sigma}
   \left[J_{1}(\frac{m_n}{k}e^\sigma)+b(m_n)
   Y_{1}(\frac{m_n}{k}e^\sigma)\right]\, ,
\end{eqnarray}
where $J_\alpha$ and $Y_\alpha$ are Bessel functions, 
$N_n$ are normalization constants  and 
the coefficients $b(m_n)$ satisfy
\begin{eqnarray}
  b(m_n)&=&-\frac{J_1(\frac{m_n}{k})}
             {Y_1(\frac{m_n}{k})}~,\\
  b(m_n)&=&b(m_n e^{\pi kR})\label{masscond}~.
\end{eqnarray}
The Kaluza-Klein masses of the gravitinos $\widetilde\psi_\mu^{(n)}$
can be obtained by solving (\ref{masscond}), and
for $n>0$ they are approximately given by
\begin{equation}
\label{kkgrav}
     m_n\simeq \left(n+\frac{1}{4}\right) \pi k e^{-\pi kR}~.
\end{equation}
Finally, using (\ref{gravi1}) and (\ref{gravi2}), 
and the fact that $\Psi_{5\, R}$
($\Psi_{5\, L}$) are  even (odd) under the 
$Z_2$-parity,
we have from the condition (\ref{condi})
\begin{eqnarray}
   f^{(n)}_{{5\, L}}&=&e^\sigma f^{(n)}_R~,\\
   f^{(n)}_{{5\, R}}&=&e^\sigma f^{(n)}_L-\frac{2\sigma^\prime}{m_n}f^{(n)}_R~.
\end{eqnarray}

\subsubsection{Massless sector}
\label{masslessmode}

The $y$-dependence of the gravitino zero-mode wavefunction is obtained from 
Eq.~(\ref{gravi1}), since under the orbifold symmetry, $f_R^{(0)}$ is 
projected out. Thus for the remaining mode, $f_L^{(0)}$ with $m_0=0$,
we obtain 
\begin{equation}
    f_L^{(0)}(y) = \frac{1}{\sqrt N_0} e^{-\frac{1}{2}\sigma}~,
\label{gravit}
\end{equation}
where the normalization factor $N_0 = (1-e^{-2\pi kR})/k$.
This is consistent with the $y$-dependence of the graviton 
zero-mode wavefunction, as expected from supersymmetry.
Similarly, $\eta_R^{(0)}$ is projected out and $\eta_L^{(0)}$ 
whose wavefunction is also given by
Eq.~(\ref{gravit})
parametrizes the remaining $N=1$ supersymmetry of the theory.
In fact one can check that Eq.~(\ref{gravit})
satisfies the Killing spinor condition.

Similarly, for the fifth-component of the gravitino, 
we have that $\psi_{5\, L}^{(0)}$ is projected out and
only $\psi_{5\, R}^{(0)}$ remains in the theory. 
This corresponds to the supersymmetric partner
of the radion, the ``radino''.
The 4d effective Lagrangian of  this field has been
recently presented in Ref.~\cite{radino}.

\section{Supersymmetry breaking  in a slice of AdS}

Different mechanisms of supersymmetry
breaking in brane-world scenarios have been considered 
in the past. The most popular, based on the 
Horava-Witten model,
corresponds to breaking supersymmetry in 
a hidden-sector living on a brane located at a finite distance 
from the observable-sector brane \cite{review}.
The moduli (e.g. the dilaton and radion) play the role
of messengers communicating the supersymmetry-breaking 
from the hidden to the observable sector.
These scenarios rely on gaugino condensation to occur
on the hidden-sector brane in order to explain the hierarchy. 

Warped (AdS) spaces allow for new possibilities. First of all,
since the hierarchy is now explained by the warp factor,
one does not need a gaugino condensation in a hidden sector to be
responsible for a small supersymmetry breaking. 
Supersymmetry can be broken at tree-level 
if it occurs on the TeV-brane, and therefore have a stringy origin. 

The supersymmetry-breaking mechanism that we will consider here
is based on imposing different boundary conditions between fermions
and bosons on the TeV-brane. This breaks supersymmetry
for the bulk fields, and as we shall see, the Kaluza-Klein
fermions and bosons
receive TeV mass-splittings.
The mechanism consists of the following. The 5d bulk fields
in the  supersymmetric $Z_2$ orbifold can be classified as either
odd or even fields under the $Z_2$ parity.
For the 5d fermions, we have two possibilities to define the $Z_2$ parity,
namely
\begin{equation}
     \psi(-y)=\pm\gamma_5\psi(y)\, .
\label{chir}
\end{equation}
Once a choice is made, 
this also defines the chirality on the 4d boundary at $y^\ast=0$, since
 $\psi(0)=\pm\gamma_5\psi(0)$.
For the supersymmetric $Z_2$ orbifold the same chirality is chosen on
the two boundaries at $y^\ast =0$ and $y^\ast=\pi R$.
In this way only half of the bulk supersymmetry is broken by the 
boundaries, leaving an $N=1$ supersymmetric theory at the massless level.
However, there also exists the possibility to separately define
the chirality of fermions on the two boundaries. For example, the choice
\begin{eqnarray}
   \psi(0)&=&\gamma_5\psi(0)\, ,\nonumber\\
   \psi(\pi R)&=&-\gamma_5\psi(\pi R)\, ,
\label{chirt}
\end{eqnarray}
corresponds to the following $y$-dependence
\begin{eqnarray}
    \psi(-y)&=&\gamma_5\psi(y)\, ,\nonumber\\
    \psi(-y+\pi R)&=&-\gamma_5\psi(y+\pi R)\, .
\label{chir2}
\end{eqnarray}
Thus, under a $2\pi$ rotation around the circle $S^1$, 
Eq.~(\ref{chir2}) leads to fermions that are antiperiodic 
\begin{equation}
\psi(y+2\pi R)=-\psi(y)\, .
\label{sss}
\end{equation}
As will be shown below, 
the boundary conditions (\ref{chirt}) 
project out the  massless fermion modes arising from  bulk fields.
Also supersymmetry is now completely broken
since no Killing spinor can be defined
\footnote{This has some similarities with finite 
temperature which also breaks supersymmetry~\cite{Brevik:2000vt}.}.

The fermionic boundary conditions (\ref{chirt}) 
have been considered previously 
in the literature. If the space of the extra dimension is flat,
imposing these boundary conditions exactly corresponds to breaking
supersymmetry by the Scherk-Schwarz mechanism \cite{ss}.
This mechanism has been applied to the MSSM
in Refs.~\cite{anto,bulk,pq,massesSS,bhn}.
In the Horava-Witten theory this was studied in Ref.~\cite{aq}.
However, for warped spaces the 
fermionic boundary conditions Eq.~(\ref{chirt}) 
do not correspond to the Scherk-Schwarz mechanism, because this  requires
a smooth limit where supersymmetry is restored \cite{ss}. 
In the case of warped spaces, we have not found such a smooth 
limit~\footnote{
Similarly, one can also show that the models of 
Refs.~\cite{pq,massesSS} with only 
one Higgs hypermultiplet (instead of two)  
do not have this smooth limit to a supersymmetric theory.
This alternative has recently been considered in 
Ref.~\cite{bhn}, where it was shown that 
the  boundary conditions Eq.~(\ref{chirt})
can also be  understood as compactifying on a 
$S^1/(Z_2\times Z_2)$ orbifold.}.

Let us now study the fermionic spectrum with the 
boundary conditions (\ref{chirt}), which we
will refer to as ``twisted'' boundary conditions.
The resulting Kaluza-Klein mass spectrum 
for $\psi_L$ 
is now determined by (see Appendix)
\begin{equation}
\label{tbc}
    \frac{J_{\alpha-1}(\frac{m_n}{k})}
      {Y_{\alpha-1}(\frac{m_n}{k})}
   = \frac{J_\alpha(\frac{m_n}{k} e^{\pi kR})}
      {Y_\alpha(\frac{m_n}{k} e^{\pi kR})}~,
\end{equation}
where $\alpha=|c+1/2|$ for a fermion of 
Dirac-type mass $m=c\sigma^\prime$
and $\alpha=2$ for the gravitino.
One can easily check that
the equation resulting from imposing twisted boundary conditions
on $\psi_R$ leads to an identical Kaluza-Klein spectrum.
The first thing to notice in Eq.~(\ref{tbc}) is that $m_n=0$ 
is no longer a solution of the above
equation and therefore no massless fermions are present.
Consequently, supersymmetry is now completely broken. 
One can also see that 
the Killing spinor, $\eta_L^{(0)}$, whose wavefunction is identical
to Eq.~(\ref{gravit}), is not consistent  with the
new boundary conditions.
Thus, the only change with respect to Section~\ref{superH}
is that Eq.~(\ref{masscond}) 
is now replaced by Eq.~(\ref{tbc}),
and the massless sector of subsection~\ref{masslessmode}, is no longer
present in the theory.
Notice that from Eq.~(\ref{varL}) the Goldstino 
of the broken $N=1$  supersymmetry is $\psi^{(0)}_{5\, L}$.

In the limit $m_n\ll k$ and $kR\gg 1$, the solution of Eq.~(\ref{tbc})
is given by
\begin{equation}
\label{kkmass}
      m_n \simeq \left(n+\frac{\alpha}{2} -\frac{1}{4}\right)\pi k 
        e^{-\pi kR}~.
\end{equation}
Comparing with the result for ``untwisted'' boundary conditions~\cite{gp},
one finds that the Kaluza-Klein mass spectrum is shifted 
by a value that
asymptotically approaches $1/2 (\pi k e^{-\pi kR})$. This is to be 
contrasted with the flat case where the shift in the Kaluza-Klein 
mass spectrum is $1/(2R)$.

There is an important difference when this type of supersymmetry
breaking is realized in warped spaces compared to the flat case.
In flat spaces this type of supersymmetry breaking is {\it global}.
To see this, let us consider an observer living on one of the branes
with the other brane sent to
infinity ($R\rightarrow\infty$). In this limit and in flat space,
supersymmetry is restored because 
the Kaluza-Klein spectrum becomes continuous
(the scalar-fermion mass splitting disappears).
This is related to the fact that one can {\it locally} ({\it i.e.},
on either brane)
define a 
supersymmetric theory.
Breaking supersymmetry globally 
(when the extra dimension is compact, no Killing spinor
can be defined in the whole space)
leads to the important property that the vacuum energy and 
the one-loop scalar masses are 
finite and independent of the cut-off scale \cite{addSS,massesSS}.

In warped spaces the situation is different and  the finiteness
of the one-loop scalar masses depends on which particular brane the
observable sector lives.
Consider first the observable sector on the Planck-brane where
the TeV-brane is sent off to infinity. In this limit supersymmetry
is restored on the Planck-brane because the Kaluza-Klein spectrum
becomes continuous. Therefore the 
one-loop scalar masses on the Planck-brane will be finite.
Alternatively, suppose that the observable sector is on the
TeV-brane. Now, even if we consider the limit where we 
send the Planck brane away ($R\rightarrow\infty$),
the Kaluza-Klein spectrum remains discrete and supersymmetry stays broken.
Therefore on the TeV-brane supersymmetry is broken and 
corrections to scalar masses
will be sensitive to the ultraviolet cut-off. 
Another way to see that 
supersymmetry is broken by the TeV-brane
(contrary to the flat case) is that no Killing spinor
can be defined 
if fermions
have twisted boundary conditions. 
Even in a non-compact space, 
the TeV-boundary breaks all the supersymmetries.
These expectations will be confirmed in the following sections
by the explicit calculation of the one-loop scalar masses
in a warped AdS space. 

Finally, as an alternative to the supersymmetry-breaking mechanism
 considered above, there also exists the possibility 
of breaking supersymmetry by the $F$-term, $F_T$ of the radion field $T$.
This can easily be achieved by turning on a constant
term, $W$, in the superpotential localized on the TeV-brane.
In flat space this is known to generate a vacuum expectation
value (VEV) $\langle F_T\rangle\sim W/M^3_5$.
In fact in flat space, this corresponds exactly to
the Scherk-Schwarz mechanism \cite{Ferrara:1994kg} or to 
imposing the twisted 
boundary condition for the fermion as in Eq.~(\ref{chirt}).
However, in a warped space this is not the case,
and a nonzero $\langle F_T\rangle$ 
leads to a new way of breaking supersymmetry.
Furthermore, in a warped space 
the VEV of  $F_T$, induced by a constant term in the superpotential
at the TeV-brane, 
is exponentially suppressed, 
$\langle F_T\rangle\sim e^{-\pi kR}W/M^3_5$.
The tree-level spectrum is easily derived. For the  gaugino we have
$m_\lambda\sim \langle F_T/T\rangle\sim$ TeV, while for scalars
localized on either brane their masses are zero.
The scalar masses are, however, induced at the one-loop level.
This scenario leads, qualitatively, to  
the same mass spectrum as the one considered above,
and will not be pursued here.

\section{The Warped MSSM}

Let us now present a candidate MSSM based on the 5d model
described above.
We will assume that both gravity and 
gauge fields are in the bulk. Supersymmetry is spontaneously broken
by imposing twisted boundary conditions, (\ref{chirt}),
  on the gravitino and gaugino.
The MSSM matter fields are assumed to be completely
localized on the Planck brane. 
At tree level, the matter fields
are massless and the dominant supersymmetry-breaking 
effects will be transmitted to the matter fields on the Planck brane by 
the 5d gauge interactions. Thus, the soft masses on the Planck brane will 
arise via radiative corrections and can be computed using the
5d AdS propagators. On the other hand, 
the Higgs field can also be assumed to
be a bulk field. However, in this case 
we will see that radiative corrections to
the Higgs soft masses are sensitive to TeV-scale physics.
Other alternative scenarios will also be discussed. Since the bulk fields
live in a warped space, 
these models will be referred to as the ``warped MSSM''.

Before proceeding to calculate the sparticle spectrum 
an important comment is in order. Since the gauge bosons live in a
warped extra dimension,
the effective 4d coupling is given by 
$g^2=(g_5^2k)/(\pi kR)$, 
where $g_5$ is the 5d gauge coupling~\cite{dhr,gauge}. 
In order to explain the Planck-TeV scale hierarchy we need
$k R\simeq 10$, which implies that for $g^2_5k\lappeq 1$, we obtain
$g^2\lappeq 1/30$. This is in 
 contradiction with the experimental values
of the gauge couplings  which require $g^2\sim {\cal O}(1)$. 
Therefore, in order to agree with the experimental values one 
requires that $g^2_5k\gappeq 30$. This inevitably means that  
the theory is close to the strong coupling regime at 
energies $E\sim k$.
On the TeV-brane this corresponds to energies $E\sim ke^{-\pi kR}$.
At these energies the expansion parameter becomes 
$g^2_5k/(16\pi^2)\sim 0.2$.
We will assume that the effects from the strong coupling regime
do not spoil the  AdS geometry. Similarly, we will  be able to 
trust our low-energy predictions, provided that the energy of the
processes satisfies $E\lappeq ke^{-\pi kR}$.

\subsection{Tree-level  masses}

If twisted boundary conditions are imposed on the  
the fermions in the bulk, then all the 4d fermion modes will receive
masses. In particular, the zero mode of the  gravitino  will receive
a mass whose magnitude can easily be obtained 
by solving (\ref{tbc}) for $\alpha=2$:
\begin{equation}
\label{gravitinomass}
      m_{3/2} \simeq \sqrt{8} k e^{-2\pi kR}~.
\end{equation}
Thus, for $k=M_P$ and $k e^{-\pi kR} =\rm TeV$ we obtain 
$m_{3/2}\simeq 2.8\times 10^{-3}{\rm eV}$. This is a superlight
gravitino, as compared to the usual gravity-mediated and gauge-mediated
scenarios in four dimensions, and satisfies the usual experimental 
constraints from cosmology and collider experiments~\cite{grav}. 
In the warped case the small 
gravitino mass arises because the coupling of the
gravitino to the TeV-brane is exponentially suppressed, and therefore it
is very insensitive to the twisting of boundary conditions on
the TeV-brane.
The higher Kaluza-Klein gravitino modes are approximately
given by 
\begin{equation}
    m_n \simeq \left(n+\frac{3}{4}\right)\pi k e^{-\pi kR}~. 
\end{equation}
Notice that compared
to the untwisted gravitino Kaluza-Klein spectrum (\ref{kkgrav}), the 
twisted mass spectrum has indeed shifted approximately by an amount 
$1/2(\pi k e^{-\pi kR})$. This shift is much larger than for the zero mode,
because the nonzero Kaluza-Klein gravitino modes are localized
near the TeV-brane and therefore couple more strongly to the
TeV brane as compared to the
the gravitino zero mode which is localized near the Planck brane.

Similarly, the tree-level gaugino mass is obtained by
solving (\ref{tbc}) with
$\alpha=1$:
\begin{equation}
\label{gauginomass}
        m_\lambda \simeq \sqrt{\frac{2}{\pi kR}}\,
        k e^{-\pi kR}~.
\end{equation}
Thus, for $k=M_P$ and $k e^{-\pi kR} =\rm TeV$ we obtain 
$m_\lambda \simeq 0.24$ TeV.
Notice that unlike the gravitino zero mode, the gaugino zero mode
receives a TeV-scale supersymmetry breaking mass. This is 
because the vector superfield is not localized in the AdS space,
and therefore directly couples to the TeV-brane, which is the source 
of the supersymmetry breaking. Using (\ref{kkmass}), the 
higher Kaluza-Klein modes are approximately given by 
$m_n \simeq (n+1/4)\pi k e^{-\pi kR}$.
These masses are obtained at tree-level and we will see that interactions
of boundary fields with the bulk gauge bosons will generate boundary masses
at one-loop. Since the mediation of the supersymmetry breaking is due to 
gauge interactions, the flavor problem is naturally solved.
It is important to note that the theory has a U(1)$_R$ symmetry,
since the induced masses are of the Dirac-type instead of the 
Majorana-type. This is a unique property of these theories, and
is due to the $N=2$ bulk supersymmetry.

It is also possible to add hypermultiplets in the bulk, where the fermions
have twisted boundary conditions. In particular, if $c=1/2$ then the
hypermultiplet is conformal, and the resulting Kaluza-Klein spectrum 
is identical to the vector supermultiplet case.

\subsection{ Radiative corrections on the Planck brane }

In order to compute the radiative corrections of the matter fields 
completely confined on the Planck brane, let us consider the 5d AdS
propagator for the gauge boson and gaugino. The general
expression for the 5d propagator in a slice of AdS
is derived in the Appendix. Using the expression for the
vector field Green's function
restricted to the Planck brane ($z=z^\prime=1/k$),  we have
\begin{equation}
\label{vectorgf}
      G_V(x,\frac{1}{k}; x^\prime,\frac{1}{k}) = \int \frac{d^4 p}{(2\pi)^4}\, 
       e^{i p\cdot(x-x^\prime)} \frac{1}{ip} \left(
       \frac{ J_0 (ip e^{\pi kR}/k)
       Y_1(ip/k)- Y_0(ipe^{\pi kR}/k) J_1(ip/k)}
       {J_0(ipe^{\pi kR}/k) Y_0(ip/k)- Y_0(ip e^{\pi kR}/k) 
       J_0(ip/k)}\right)~.
\end{equation}
In the limit that $p\ll ke^{-\pi kR}$ we obtain
\begin{equation}
\label{planckGs}
      G_V(x,\frac{1}{k}; x^\prime,\frac{1}{k}) \simeq
       \frac{-1}{\pi R} \int \frac{d^4 p}{(2\pi)^4}
       e^{i p\cdot(x-x^\prime)}\, \frac{1}{p^2}~, 
\end{equation}
which reduces to the usual massless vector field 
Green's function in flat space.
In particular notice that by Eq.~(\ref{planckGs}), the 
charge screening~\cite{log,Kaloper:2000xa} 
is absent in the slice of AdS since 
there are no continuum Kaluza-Klein modes.
Similarly on the Planck-brane, the twisted gaugino Green's function
defined in the Appendix reduces to the form
\begin{equation}
\label{twistedf}
      G_F(x,\frac{1}{k}; x^\prime,\frac{1}{k}) = \int \frac{d^4 p}{(2\pi)^4}\, 
       e^{i p\cdot(x-x^\prime)} \frac{1}{ip} \left(
       \frac{ J_1 (ip e^{\pi kR}/k)
       Y_1(ip/k)- Y_1(ip e^{\pi kR}/k) J_1(ip/k)}
       {J_1(ip e^{\pi kR}/k) Y_0(ip/k)- Y_1(ip e^{\pi kR}/k) 
       J_0(ip/k)}\right)~.
\end{equation}
In the limit that $p\ll ke^{-\pi kR}$ and $kR\gg 1$
the twisted gaugino Green's function becomes
\begin{equation}
      G_F(x,\frac{1}{k}; x^\prime,\frac{1}{k}) \simeq
       \frac{-1}{\pi R} \int \frac{d^4 p}{(2\pi)^4}
       e^{i p\cdot(x-x^\prime)}\, \frac{1}{p^2-\frac{2}{\pi k R}
        (k e^{-\pi k R})^2}~, 
\end{equation}
which reduces to a massive gaugino Green's function where the gaugino
mass agrees with (\ref{gauginomass}). This difference
between the gauge boson and gaugino Green's function
represents the source of supersymmetry breaking on the Planck brane.

Note that the vector supermultiplet in the bulk,
is also equivalent to a conformal hypermultiplet ($c=1/2$)
in the bulk, where the fermion has twisted boundary conditions. 
The 5-dimensional mass-squared of the scalar is 
$-3 k^2 + 2k (\delta(y)-\delta(y-\pi R))$, while the
mass of the fermion is $\sigma^\prime/2$~\cite{gp}. 
On the Planck brane $(z=z^\prime=1/k)$
the twisted fermion Green's function is the same as Eq.~(\ref{twistedf}),
while the scalar field Green's function is identical
to Eq.~(\ref{vectorgf}).
In particular, we will also consider the bulk Higgs fields to be 
conformal hypermultiplets.

The scalar and twisted fermion Green's function on the Planck brane
can be used to calculate the one-loop contribution to the mass-squared 
of boundary matter fields.
The boundary matter fields couple to the vector supermultiplet in the bulk 
via gauge interactions. The Feynman diagrams for the 
one-loop mass contributions to the
boundary scalar fields are the same as those in flat space and
can be found in Ref.~\cite{massesSS}.
They  give
\begin{equation}
\label{mass2}
     m_i^2 = 4 g^2 C(R_i) \Pi(0)~,
\end{equation}
where the 4d gauge coupling is given by $g^2=g_5^2/(\pi R)$ and 
we have defined
\begin{equation}
\label{pidefn}
     \Pi(0)=-\pi R\int \frac{d^4 p}{(2\pi)^4}\,
     \left[ G^{(V)}_p - G^{(F)}_p\right]~.
\end{equation}
The coefficient $C(R_i)$ is the quadratic Casimir of the 
representation $R_i$ in the
corresponding gauge group, and $G_p$ is the 4d Fourier transform
of the Green's function (see Appendix).
Similarly, the boundary matter fields can 
also couple to a chiral supermultiplet
in the bulk \cite{massesSS}. 
For a conformal supermultiplet ($c=1/2$), where the
fermions have twisted boundary conditions we obtain
\begin{equation}
     m_i^2 =  Y^2 \Pi(0)~,
\end{equation}
where $Y$ is the boundary-bulk Yukawa coupling.
Assuming that $k = M_P$ and $k e^{-\pi kR}= \rm TeV$ then we obtain
$\pi k R = 34.54$, and the integral in Eq.~(\ref{pidefn}) can be
numerically evaluated to give
\begin{equation}
     \Pi(0)= \frac{0.2525}{2 \pi^4} ({\rm TeV})^2\simeq 
        (0.0360\,\rm TeV)^2~.
\end{equation}
This result is finite and insensitive to the ultraviolet cut-off
for the same reason that we already explained in the previous section, 
namely that the supersymmetry
breaking is localized on the TeV-brane.
In fact, the integration 
region $p\leq ke^{-\pi kR}=$ TeV already contributes 
approximately 90$\%$ of the integral in Eq.~(\ref{pidefn}).

Comparing the above result with the flat space case 
where~\cite{massesSS} $\Pi(0)= (0.0367/R_{flat})^2$,
we see that the two cases  are almost numerically identical for a  flat-space
radius of $R_{flat}=k e^{-\pi k R}$.

\subsubsection{Superparticle spectrum}

It is now straightforward to extend the above result to the case
of the warped MSSM. Assuming that the squarks and sleptons live
on the Planck-brane they will receive a one-loop contribution
from the bulk gauge and Higgs sector (if they are in the bulk). 
Assuming the bulk Higgs
to be a conformal supermultiplet and following Ref.~\cite{massesSS} 
we obtain
\begin{eqnarray}
      m_{\widetilde Q}^2 &=& \left(\frac{4}{3} \alpha_3 
       + \frac{3}{4}\alpha_2+\frac{1}{60}\alpha_1 \right) \Pi(0)
       +\frac{1}{2}(\alpha_t+\alpha_b) \Pi(0)~, \\
      m_{\widetilde U}^2 &=& \left(\frac{4}{3} \alpha_3 
       + \frac{4}{15}\alpha_1 \right) \Pi(0)
       +\alpha_t \Pi(0)~, \\
      m_{\widetilde D}^2 &=& \left(\frac{4}{3} \alpha_3 
       + \frac{1}{15}\alpha_1 \right) \Pi(0)
       +\alpha_b \Pi(0)~, \\
      m_{\widetilde L}^2 &=& \left(\frac{3}{4} \alpha_2 
       + \frac{3}{20}\alpha_1 \right) \Pi(0)
       +\alpha_\tau \Pi(0)~, \\
      m_{\widetilde E}^2 &=& \frac{3}{5} \alpha_1 \Pi(0)~,
\end{eqnarray}
where the bulk gauge contribution is proportional to the gauge
couplings $\alpha_{1,2,3}$ and the conformal bulk Higgs contribution is
proportional to the Yukawa couplings $\alpha_{t,b,\tau}$.
Thus to obtain an experimentally allowed soft mass spectrum the scale 
on the TeV brane should be at least a few TeV. Notice that the 
dominant corrections are proportional to the gauge couplings. Thus, the
lightest scalar field is the right-handed slepton.

\subsection{Radiative corrections in the slice of AdS}

Consider a conformal hypermultiplet in the bulk with
twisted boundary conditions for the fermion.
The massless scalar mode $\phi$ in the hypermultiplet will receive
a one-loop mass contribution due to the breaking of supersymmetry from
fields in the bulk with the twisted boundary conditions. 
In particular the scalar can couple to the bulk vector supermultiplet.
This radiative correction can simply
be calculated using the 5-dimensional AdS scalar propagator.
For an alternative method to calculate  quantum effects in
the   AdS slice see Ref.~\cite{gpt}.
Since the scalar propagates in the bulk we need to integrate 
over the extra dimension,
and the corresponding mass correction is proportional to
\begin{equation}
\label{bulkcorr}
       \Pi(0) = -\pi R\int \frac{d^4 p}{(2\pi)^4}\,
       \int_0^{\pi R} dy \, \left[G_p^{(V)}(z,z) 
            -G_p^{(F)}(z,z)\right] 
        \simeq \frac{\Lambda k\, e^{-2\pi kR}}{16\pi^2}
\end{equation}
where $\Lambda$ is a Planck-scale cutoff. 
Unlike the radiative corrections
of the boundary fields, it turns out that the radiative correction 
(\ref{bulkcorr}) is not finite, as expected from the arguments
of the previous section. 
In fact, the bulk radiative 
corrections (\ref{bulkcorr}) are linearly divergent. This reflects the
fact that the bulk fields are propagating in five dimensions
and are sensitive to physics on the TeV-brane represented by the cutoff
scale $\Lambda e^{-\pi kR}$. 
This behavior is related to the fact that
the supersymmetry breaking mechanism is localized on the TeV-brane
and is sensitive to the UV physics. This is different from flat space
where the breaking of supersymmetry is inherently a global effect,
and consequently the nonlocality produces a finite result.

Let us finally comment on other possible alternatives.
If the Higgs is also confined on the Planck-brane, then 
its mass will be generated at the one-loop level, with a magnitude
similar to that of the sleptons (without the Yukawa coupling contribution).
Although this contribution will be positive, there are sizeable 
two-loop effects arising from the squarks that 
can make the mass-squared negative~\cite{massesSS}.
Unfortunately, the Higgsino mass cannot be generated by radiative
corrections and we will need to extend the model to include
a Higgs singlet whose VEV must induce the Higgsino mass. 
In the above cases we have restricted
the observable sector to the Planck-brane. Nevertheless,
many more possibilities
exist by placing part of the matter in the bulk or on the TeV-brane.
For example, consider delocalizing the first two families off 
the Planck-brane by changing their bulk mass parameters~\cite{gp}.
In this case the corresponding  squarks and sleptons of the 
first two families will 
have masses larger that those of the  third family, a scenario 
whose phenomenology can have interesting consequences.

\subsection{Relation to 4d strongly-coupled CFT}

The AdS/CFT correspondence relates the 5d theory of gravity 
in AdS to a 4d strongly coupled conformal field theory 
(CFT)~\cite{maldacena}.
In the case of a slice of AdS, a similar correspondence
can also be formulated~\cite{gu,gkr,apr,rz}.
The Planck-brane in AdS$_5$ 
corresponds to an ultraviolet cutoff of the 4d CFT
and to the gauging of certain global symmetries. For example, in the 
case we are considering where gravity and the standard model 
gauge bosons live in the bulk,
the corresponding CFT will have the superPoincare group gauged
(giving rise to gravity) and also the standard model 
group SU(3)$\times$SU(2)$_L\times$U(1)$_Y$
(giving rise to the standard model gauge bosons and gauginos).
Matter on the Planck-brane corresponds to adding new fields 
to the CFT which only couple to CFT states via
gravity and gauge interactions.
On the other hand, the TeV-brane corresponds in the dual theory 
to a infrared cutoff of the CFT \cite{apr,rz}.
In other words, it corresponds to breaking 
the conformal symmetry at the TeV-scale.
The Kaluza-Klein states of the 5d theory correspond to the bound 
states of the strongly coupled CFT.

This alternate dual description suggests that the supersymmetry-breaking 
mechanism that we have discussed
represents a class of strongly coupled CFT's where 
supersymmetry is broken at the TeV scale. The bound states therefore 
do not respect supersymmetry and give rise to a fermion-boson mass
splitting.
The 5d warped MSSM is then simply the ordinary 4d MSSM with a 
strongly coupled CFT sector responsible for the breaking of supersymmetry.
The standard model fields coupled to the CFT sector will get tree-level 
masses while those coupled only via gravity or gauge interactions
will receive masses at the one-loop level.
In our model the CFT sector is charged under the standard 
model gauge group and consequently it 
implies that the gauginos get masses at tree-level.
Notice that as we mentioned earlier the gaugino mass is of the 
Dirac-type. This means that the gaugino has married a fermion 
bound-state to become massive~\footnote{A Majorana-type mass would correspond, 
for example, to a breaking of supersymmetry (in the 5d dual) 
by a nonzero $F_T$.}.
Since the  gaugino mass comes from the mixing between the gaugino
and the CFT bound-state, the mass will be proportional to
$\sqrt{g^2 b_{CFT}/(8\pi^2)}=1/\sqrt{\pi k R}$,
where we have used the AdS/CFT relation \cite{apr} 
$g^2_5k=8\pi^2/b_{CFT}$ and 
$g^2=g_5^2/(\pi R)$.
This agrees with Eq.~(\ref{gauginomass}).
Similarly, 
the smallness of the  gravitino mass
(of order $10^{-3}$ eV) is also easy to  understand in the 
CFT picture. 
The gravitino coupling to the CFT sector is suppressed 
by $1/M_P$, so its mass will be of order TeV$^2/M_P\sim 10^{-3}$ eV.

Although the CFT picture is useful for understanding some qualitative 
aspects of the theory, it is practically useless for obtaining
quantitative predictions since the theory is strongly coupled.
In this sense, the 5d gravitational theory in a slice of AdS represents
a very useful tool since it allows one to calculate the particle 
spectrum, which would otherwise be unknown from the CFT side.

\section{Conclusion}

In this paper we have presented a supersymmetric 5d theory in warped space
where supersymmetry is  spontaneously broken by imposing different  boundary
conditions between the fermion and bosons.
While this is reminiscent of the
Scherk-Schwarz mechanism in flat space, we have argued that in a warped space
this is a novel way of breaking supersymmetry.
Unlike the flat-space case where the supersymmetry-breaking mechanism is a
global effect, the twisted boundary conditions in the 
warped space lead to
a local supersymmetry breaking effect on the TeV-brane.

A particularly interesting model is the warped MSSM, 
where matter is  confined on the Planck brane, and gravity 
and gauge fields propagate in the 5d bulk. 
The gravitino and gaugino receive tree-level masses from the  
twisted boundary conditions.
In particular, the tree-level mass of the gravitino is $\sim 10^{-3}$ eV
and the gaugino mass $\sim$ TeV. The one-loop radiative corrections to the
squarks and sleptons confined to the Planck brane are finite and
insensitive to the UV cutoff.  This simply reflects the fact that the
supersymmetry-breaking is localized on the TeV-brane, at a finite distance
away from the Planck-brane.  The one-loop radiative corrections from the
bulk gauge fields are proportional to the gauge couplings and thus
naturally solve the flavor problem. If the Higgs sector is also included
in the bulk, then the one-loop radiative corrections also give a
contribution proportional to the Yukawa couplings. However, in this case
the radiative corrections to the Higgs soft 
mass are not finite. This is in
contrast to the flat-space case, and is due to the fact that the bulk
Higgs directly couples to the supersymmetry breaking effects on the
TeV-brane.

By the AdS/CFT correspondence, the warped supersymmetric standard model
can be interpreted in terms of a strongly coupled CFT, where supersymmetry
(and conformal symmetry) are broken at the TeV-scale. Thus, the warped
MSSM is simply the ordinary 4d MSSM with a strongly coupled CFT which is
responsible for breaking supersymmetry. The fact that there exists a
weakly coupled 5d gravity dual, allows us to calculate the mass spectrum.
This provides a powerful tool in obtaining information about the dynamics
of this class of strongly coupled CFT's, and is worthy of further
investigation.

\section*{Acknowledgements}
We wish to thank Emilian Dudas and Dan Waldram for helpful discussions. 
One of us (TG) acknowledges the Aspen Center for Physics where
part of this work was done.
The work of TG is supported by the FNRS, contract no. 21-55560.98, 
while that of AP is partially supported by the CICYT Research Project
AEN99-0766.

\section*{Appendix: 5D propagators in a slice of AdS}

Let us consider the propagation of bulk fields in a slice of AdS.
We will follow the derivation of the Green's function presented 
in Ref.~\cite{gt,gkr}, except that we will extend the previous results 
to the case of arbitrary bulk fields in the two-brane scenario. The
result for the bulk scalar has also recently been given in 
Ref.~\cite{gns}.  

As shown in Ref.~\cite{gp} the equation of motion for bulk fields
$\Phi=\{V_\mu, \phi, e^{-2\sigma} \psi_{L,R}\}$, 
can be conveniently written as a second-order differential
equation. Thus, introducing a source function $\cal J$,
one obtains 
\begin{equation}
\label{geneqn}
     \left[e^{2\sigma} \eta^{\mu\nu} \partial_\mu \partial_\nu 
       + e^{s\sigma}\partial_5(e^{-s\sigma}\partial_5)
       -M_\Phi^2\right] \Phi(x,y)= {\cal J}(x,y)~,
\end{equation}
where the parameter $s=\{2,4,1\}$, and
the 5d masses are  \cite{gp} $M_\Phi^2=\{0,ak^2+b\sigma^{\prime\prime},
c(c\pm 1)k^2\mp c \sigma^{\prime\prime}\}$.
The corresponding Green's function for (\ref{geneqn}) 
can then be defined as
\begin{equation}
    \Phi(x,y) = \int d^4 x^\prime dy^\prime \sqrt{-g} \,
       G(x,y; x^\prime,y^\prime) e^{(4-s)k y^\prime}{\cal J}
      (x^\prime,y^\prime)~.
\end{equation}
provided that ${\cal J}=\{J_\mu, J_\phi, \gamma^\mu \partial_\mu 
J_{R,L}\pm\partial_5 J_{L,R}-(c\pm 1) \sigma^\prime 
J_{L,R}\}$, where $J_\mu$, $J_\phi$ and $J_{L,R}$
are the source terms for the bulk vector, scalar, and fermion, respectively.
It is convenient to introduce the variable $z=e^{ky}/k$. In these
coordinates the Planck-brane is located at $z^\ast = 1/k$ and the
TeV-brane at $z^\ast = e^{\pi kR}/k$. If we now
take the 4d Fourier transform of the Green's function
\begin{equation}
     G(x,z;x^\prime,z^\prime) = \int \frac{d^4 p}{(2\pi)^4} 
      e^{ip\cdot (x-x^\prime)} G_p(z,z^\prime)~,
\end{equation}
then the Fourier component $G_p(z,z^\prime)$ must satisfy the equation
\begin{equation}
\label{Gpeqn}
    \left(\partial_z^2 + \frac{1-s}{z}\partial_z - p^2 - 
      \frac{{\widehat M}_\Phi^2}{(kz)^2} \right) G_p(z,z^\prime)
      = (kz)^{s-1} \delta(z-z^\prime)~,
\end{equation}
where ${\widehat M}_\Phi^2=\{0,ak^2,c(c\pm 1)k^2 \}$. If we define  
$G_p(z,z^\prime) = (k^2 z z^\prime)^{s/2} {\widehat G}_p(z,z^\prime)$, then
Eq.~({\ref{Gpeqn}) simply becomes the Bessel equation
\begin{equation}
\label{greenfneqn}
    \left(\partial_z^2 + \frac{1}{z} \partial_z - p^2 - \frac{\alpha^2}{z^2}
      \right) {\widehat G}_p(z,z^\prime)
      = (kz)^{-1} \delta(z-z^\prime)~,
\end{equation}
where $\alpha= \sqrt{(s/2)^2 + {\widehat M}_\Phi^2/k^2}$. The standard 
procedure for solving Eq.~(\ref{greenfneqn}) is to use the solution to the
homogeneous equation in the regions $z<z^\prime$ and $z>z^\prime$,
and then impose matching conditions at $z=z^\prime$.
Thus writing
\begin{equation}
    {\widehat G}_p(z,z^\prime) = \theta(z-z^\prime) {\widehat G}_> +
           \theta(z^\prime - z) {\widehat G}_<~,
\end{equation}
the solution to the homogeneous equation for $z < z^\prime$ is given by
\begin{equation}
      {\widehat G}_<(z,z^\prime) = i A_<(z^\prime) 
       \left[ {\widetilde J}_\alpha(ip/k) H_\alpha^{(1)}(ip z) - 
              {\widetilde H}_\alpha^{(1)} (ip/k) J_\alpha(ip z) \right]~,
\end{equation}
and for $z>z^\prime$ we obtain
\begin{equation}
      {\widehat G}_>(z,z^\prime) = i A_>(z^\prime) 
     \left[ {\widetilde J}_\alpha(ipe^{\pi k R}/k) H_\alpha^{(1)}(ip z) - 
         {\widetilde H}_\alpha^{(1)} (ip e^{\pi k R}/k) 
          J_\alpha(ip z) \right]~,
\end{equation}
where $H_\alpha^{(1)} = J_\alpha+ i Y_\alpha$ is the Hankel function 
of the 1st kind of order $\alpha$, and $J_\alpha$,$Y_\alpha$ are 
Bessel functions.
If the boundary condition for the Green's function, 
${\widehat G}_< ({\widehat G}_>)$
is even on the Planck-brane (TeV-brane) then \cite{gp}
\begin{equation}
      \widetilde J_\alpha(z) = (-r+s/2) J_\alpha(z)+ zJ_\alpha^\prime(z)~,
\label{jtilde}
\end{equation}
where the parameter $r=\{0,b,\mp c\}$, 
while if the boundary condition is odd then \cite{gp}
\begin{equation}
      \widetilde J_\alpha(z) = J_\alpha(z)~,
\label{jtilde2}
\end{equation}
and similarly for $\widetilde H_\alpha^{(1)}$.
Note that in the presence of the boundary mass terms parametrized
by $r$, the even boundary condition is equivalent to  
imposing the modified Neumann condition,
$(\partial_z -r \sigma^\prime) G_p(z,z^\prime) \big |_{z = z^\ast} = 0$,
while the odd boundary condition is equivalent to imposing
the Dirichlet condition,
$G_p(z,z^\prime)\big |_{z = z^\ast} =0$.

The unknown functions $A_<(z^\prime)$
and $A_>(z^\prime)$ are determined by imposing matching conditions
at $z=z^\prime$. Continuity of $\widehat G_p$ at $z=z^\prime$ leads to
the condition
\begin{eqnarray}
  {\widehat G}_> \big |_{z=z^\prime} = {\widehat G}_< \big |_{z=z^\prime}~, 
\end{eqnarray}
while the discontinuity in $\partial_z {\widehat G}_p$ gives
the condition
\begin{eqnarray}
      \left(\partial_z {\widehat G}_>-  \partial_z {\widehat G}_< 
      \right) \bigg |_{z=z^\prime}= \frac{1}{k z^\prime}~.
\end{eqnarray}
This leads to the solutions
\begin{eqnarray}
     A_<(z^\prime) &=& \frac{\pi}{2 k} 
         \frac{{\widetilde J}_{\alpha}(ip e^{\pi kR}/k) 
         H_\alpha^{(1)}(ip z^\prime)- {\widetilde H}_{\alpha}^{(1)}
         (ip e^{\pi kR}/k) J_\alpha(ip z^\prime)}
         {{\widetilde J}_{\alpha}(ip e^{\pi kR}/k) 
         {\widetilde H}_\alpha^{(1)}(ip/k)- {\widetilde H}_{\alpha}^{(1)}
         (ip e^{\pi kR}/k) {\widetilde J}_\alpha(ip/k)}~,\\
     A_>(z^\prime)& =& \frac{\pi}{2 k} 
         \frac{{\widetilde J}_{\alpha}(ip/k ) 
         H_\alpha^{(1)}(ip z^\prime)- {\widetilde H}_{\alpha}^{(1)}
         (ip/k) J_\alpha(ip z^\prime)}
         {{\widetilde J}_{\alpha}(ip e^{\pi kR}/k) 
         {\widetilde H}_\alpha^{(1)}(ip/k)- {\widetilde H}_{\alpha}^{(1)}
         (ip e^{\pi kR}/k) {\widetilde J}_\alpha(ip/k)}~.
\end{eqnarray}
Finally substituting these functions into the equations 
for $\widehat G_>$ and $\widehat G_<$ gives the expression for 
the Green's function in a slice of AdS
\begin{eqnarray}
\label{greenfn}
     G_p(z,z^\prime)&=&i \frac{\pi}{2} k^{s-1} (z z^\prime)^{s/2}
        \left[\frac{{\widetilde J}_{\alpha}(ip e^{\pi kR}/k) 
         H_\alpha^{(1)}(ip z_>)- {\widetilde H}_{\alpha}^{(1)}
         (ip e^{\pi kR}/k) J_\alpha(ip z_>)}
       {{\widetilde J}_{\alpha}(ip e^{\pi kR}/k) 
       {\widetilde H}_\alpha^{(1)}(ip/k)- {\widetilde H}_{\alpha}^{(1)}
      (ip e^{\pi kR}/k) {\widetilde J}_\alpha(ip/k)} \right] \nonumber \\
    && \times \left[{\widetilde J}_{\alpha}(ip/k) H_\alpha^{(1)}(ip z_<)-
        {\widetilde H}_{\alpha}^{(1)}(ip/k) J_\alpha(ip z_<)\right]~,
\end{eqnarray}
where we have defined $z_> (z_<)$ to be the greater (lesser) of $z$ and 
$z^\prime$. The Green's function (\ref{greenfn}) is the general expression
for arbitrary bulk fields in a slice of AdS. 

The Kaluza-Klein mass spectrum can be obtained from the pole condition
of the Green's function, namely
\begin{equation}
\label{masscon}
    {{\widetilde J}_{\alpha}(ip e^{\pi kR}/k) 
    {\widetilde H}_\alpha^{(1)}(ip/k)- {\widetilde H}_{\alpha}^{(1)}
      (ip e^{\pi kR}/k) {\widetilde J}_\alpha(ip/k)} = 0~.
\end{equation}
This leads to the condition
\begin{equation}
\label{genbc}
     \frac{\widetilde J_\alpha (\frac{m}{k})}{\widetilde Y_\alpha 
      (\frac{m}{k})}=
     \frac{\widetilde J_\alpha (\frac{m}{k}e^{\pi kR})}{\widetilde 
     Y_\alpha (\frac{m}{k}e^{\pi kR})}~, 
\end{equation}
where the four-momentum $p^2= -m^2$. 
The solutions of this equation for the various combinations
of boundary conditions on the Planck and TeV-branes, 
reproduces all the Kaluza-Klein mass spectrum results from~\cite{gp}.

For the twisted boundary condition, $\Phi(0)=\Phi(0)$ and 
$\Phi(\pi R)=-\Phi(\pi R)$, 
we must impose on the corresponding Green's function 
the even boundary condition on the Planck-brane, Eq.~(\ref{jtilde}), 
and the odd boundary condition on the TeV-brane, Eq.~(\ref{jtilde2}).
The mass spectrum is obtained by solving (\ref{genbc}) which now 
becomes
\begin{equation}
     \frac{(-r+s/2)J_\alpha (\frac{m}{k})
+\frac{m}{k}J^\prime_\alpha (\frac{m}{k})}
{(-r+s/2)Y_\alpha(\frac{m}{k})+\frac{m}{k}Y^\prime_\alpha(\frac{m}{k})}
   =\frac{J_\alpha (\frac{m}{k}e^{\pi kR})}
{ Y_\alpha (\frac{m}{k}e^{\pi kR})}~, 
\end{equation}
and for the fermion and gravitino the equation
simplifies further to Eq.~(\ref{tbc}).

\end{document}